\documentstyle[12pt,epsf,rotate]{article}

\def\spose#1{\hbox to 0pt{#1\hss}}
\def\lsim{\mathrel{\spose{\lower 3pt\hbox{$\mathchar"218$}}
 \raise 2.0pt\hbox{$\mathchar"13C$}}}
\def\gsim{\mathrel{\spose{\lower 3pt\hbox{$\mathchar"218$}}
 \raise 2.0pt\hbox{$\mathchar"13E$}}}

\begin{document}

\begin{titlepage}

\begin{flushright}
CERN-TH/98-128\\
hep-ph/9804319 
\end{flushright}

\vspace{2cm}
\begin{center}
\boldmath
\large\bf
Controlling Rescattering Effects in Constraints on the\\ 
\vspace{0.2truecm}
CKM Angle $\gamma$ arising from $B\to\pi K$ Decays
\unboldmath
\end{center}

\vspace{1.2cm}
\begin{center}
Robert Fleischer\\[0.1cm]
{\sl Theory Division, CERN, CH-1211 Geneva 23, Switzerland}
\end{center}

\vspace{1.2cm}
\begin{abstract}
\vspace{0.2cm}\noindent
It has recently been pointed out that the observables of the decay 
$B^+\to K^+\overline{K^0}$ and its charge conjugate allow us to take 
into account rescattering effects in constraints on the CKM angle $\gamma$ 
arising from $B^\pm\to\pi^\pm K$ and $B_d\to\pi^\pm K^\pm$ modes, and that
they play an important role to obtain insights into final-state interactions. 
In this paper, the formalism needed to accomplish this task is discussed in 
detail. Furthermore, using a transparent model to describe the rescattering 
processes, as well as electroweak penguins, we calculate the quantities 
parametrizing the $B^+\to\pi^+ K^0$ and $B_d^0\to\pi^- K^+$ decay 
amplitudes for specific examples, and illustrate the constraints on 
$\gamma$ arising from the corresponding observables. Although this model 
is very crude, it shows nicely the power of $B^\pm\to K^\pm K$ both to 
include the rescattering effects in the bounds on $\gamma$ and to obtain 
insights into final-state interactions. Moreover this model exhibits the 
interesting feature that the combined branching ratio BR$(B^\pm\to K^\pm K)$ 
may be considerably enhanced through rescattering processes, as was recently 
pointed out within a general framework. 
\end{abstract}

\vfill
\noindent
CERN-TH/98-128\\
April 1998

\end{titlepage}

\thispagestyle{empty}
\vbox{}
\newpage
 
\setcounter{page}{1}

\section{Introduction}\label{intro}
The issue of rescattering effects in $B\to\pi K$ decays, originating from 
processes such as $B^+\to\{\pi^0K^+\}\to\pi^+K^0$, led to considerable 
interest in the recent literature \cite{bfm}--\cite{atso} (for earlier 
references, see \cite{FSI}). An important implication of these final-state 
interaction effects may be direct CP violation, in the mode $B^+\to\pi^+K^0$,
as large as ${\cal O}(10\%)$, whereas estimates performed at the perturbative 
quark level, following the approach proposed by Bander, Silverman and Soni 
\cite{bss}, typically give CP asymmetries of at most a few 
percent~\cite{pert-pens}. Several papers dealing with these rescattering 
effects tried to point out that they would also invalidate bounds
on the angle $\gamma$ of the usual non-squashed unitarity triangle~\cite{ut}
of the Cabibbo--Kobayashi--Maskawa matrix (CKM matrix) \cite{ckm} that were 
derived in \cite{fm2}; they would arise if the ratio
\begin{equation}\label{Def-R}
R\equiv\frac{\mbox{BR}(B_d\to\pi^\mp K^\pm)}{\mbox{BR}(B^\pm\to\pi^\pm K)}
\end{equation}
of the combined branching ratios
\begin{eqnarray}
\mbox{BR}(B_d\to\pi^\mp K^\pm)&\equiv&\frac{1}{2}
\left[\mbox{BR}(B^0_d\to\pi^-K^+)
+\mbox{BR}(\overline{B^0_d}\to\pi^+K^-)\right]\label{BR-neut}\\
\mbox{BR}(B^\pm\to\pi^\pm K)&\equiv&\frac{1}{2}\left[\mbox{BR}(B^+\to\pi^+K^0)
+\mbox{BR}(B^-\to\pi^-\overline{K^0})\right]\label{BR-char}
\end{eqnarray}
is found to be smaller than 1. These quantities have recently been measured 
for the first time by the CLEO collaboration, and the present experimental 
results are as follows \cite{cleo}: 
\begin{eqnarray}
\mbox{BR}(B_d\to\pi^\mp K^\pm)&=&\left(1.5^{+0.5}_{-0.4}\pm0.1\pm0.1
\right)\times10^{-5}\label{BR-neutres}\\
\mbox{BR}(B^\pm\to\pi^\pm K)&=&\left(2.3^{+1.1}_{-1.0}\pm0.3\pm0.2\right)
\times 10^{-5}\label{BR-charres}\,,
\end{eqnarray}
yielding $R=0.65\pm0.40$. Consequently, it may well be that future 
measurements will stabilize at a value of $R$ that is significantly 
smaller than 1, thereby leading to interesting constraints on the 
CKM angle $\gamma$ \cite{fm2} (for a detailed study, see \cite{gnps}). 
At first sight, an important limitation of the 
theoretical accuracy of these bounds is in fact due to the rescattering 
effects mentioned above \cite{gewe}--\cite{atso}. A closer look shows,
however, that these effects do not spoil the bounds on $\gamma$, and can 
be included completely, in a rather straightforward way, through the decay 
$B^+\to K^+\overline{K^0}$ and its charge conjugate, providing in addition 
valuable insights into final-state interactions \cite{th98-60}. The combined
branching ratio BR$(B^\pm\to K^\pm K)$ for this decay, which is defined in
analogy to (\ref{BR-neut}) and (\ref{BR-char}), is very sensitive to 
rescattering processes and may be enhanced considerably through them. 

In this paper, we illustrate these interesting features in more detail by 
following closely \cite{th98-60} and using the parametrization of
the $B^+\to\pi^+K^0$ and $B^0_d\to\pi^-K^+$ decay amplitudes in terms of
the ``physical'' quantities given there. The outline is as follows: in 
Section~\ref{sec:obs}, we collect the expressions for the $B\to\pi K$
decay amplitudes, introduce the relevant observables, and discuss briefly
the constraints on the CKM angle $\gamma$ implied by them. 
In Section~\ref{sec:FSIcontrol}, we focus on the approach to take into
account rescattering effects in these bounds with the help of 
$B^\pm\to K^\pm K$. This strategy, as well as the realization of the
constraints on $\gamma$, is illustrated in a quantitative way in 
Section~\ref{sec:model}, where we will use a simple model to describe the
rescattering effects that has been proposed in \cite{gewe,neubert} and is 
based on the assumption of elastic final-state interactions. In 
Section~\ref{sec:sum}, a few concluding remarks are given. 

\boldmath
\section{Decay Amplitudes, Observables and Constraints on the CKM Angle 
$\gamma$}\label{sec:obs}
\unboldmath
The general $B^+\to\pi^+K^0$ and $B^0_d\to\pi^-K^+$ decay 
amplitudes arising within the framework of the Standard Model can be 
expressed as follows \cite{th98-60}:
\begin{eqnarray}
A(B^+\to\pi^+K^0)&=&P\label{ampl-neut}\\
A(B^0_d\to\pi^-K^+)&=&-\,[P+T+P_{\rm ew}]\label{ampl-char}\,.
\end{eqnarray}
In order to derive these amplitude relations, the $SU(2)$ isospin symmetry 
of strong \mbox{interactions} has been used. The amplitude $P$, which is 
usually referred to as a ``penguin'' amplitude, takes the form
\begin{equation}\label{P-def}
P=-\left(1-\frac{\lambda^2}{2}\right)\lambda^2A\left[
1+\rho\,e^{i\theta}e^{i\gamma}\right]{\cal P}_{tc}\,,
\end{equation}
where
\begin{equation}\label{rho-def}
\rho\,e^{i\theta}=\frac{\lambda^2R_b}{1-\lambda^2/2}
\left[1-\left(\frac{{\cal P}_{uc}+{\cal A}}{{\cal P}_{tc}}\right)\right].
\end{equation}
Here the quantities ${\cal P}_{tc}\equiv|{\cal P}_{tc}|e^{i\delta_{tc}}$ and 
${\cal P}_{uc}\equiv|{\cal P}_{uc}|e^{i\delta_{uc}}$ describe contributions 
originating from penguin topologies with internal top and charm, and up and 
charm quarks, respectively, ${\cal A}$ is due to annihilation processes, and
\begin{equation}
\lambda\equiv|V_{us}|=0.22\,,\quad
A\equiv\frac{1}{\lambda^2}\left|V_{cb}\right|=0.81\pm0.06\,,\quad
R_b\equiv\frac{1}{\lambda}\left|\frac{V_{ub}}{V_{cb}}\right|=0.36\pm0.08
\end{equation}
are the relevant CKM factors, expressed in terms of the Wolfenstein 
parameters \cite{wolf}. The amplitudes $T$ and $P_{\rm ew}$ -- the latter 
is due to electroweak penguins -- can be parametrized in a simple way as
\begin{equation}\label{TPew}
T\equiv|T|\,e^{i\delta_T}e^{i\gamma}\,,\quad P_{\rm ew}\equiv-\,|P_{\rm ew}|
\,e^{i\delta_{\rm ew}}\,,
\end{equation}
where $\delta_T$ and $\delta_{\rm ew}$ are CP-conserving strong phases such
as $\delta_{tc}$, $\delta_{uc}$ and $\theta$. The expressions for the 
charge-conjugate decays $B^-\to\pi^-\overline{K^0}$ and 
$\overline{B^0_d}\to\pi^+K^-$ can be obtained straightforwardly from
(\ref{ampl-neut}) and (\ref{ampl-char}) by performing the substitution
$\gamma\to-\,\gamma$ in (\ref{P-def}) and (\ref{TPew}), i.e.\ $|T|$ and 
$|P_{\rm ew}|$, in contrast to $|P|$, exhibit no CP violation. In the 
literature, $T$ is usually referred to as a ``tree'' amplitude. This 
terminology is, however, misleading in this case, since $T$ actually
receives not only ``tree'' contributions, but also contributions from 
penguin and annihilation topologies, as was pointed out in \cite{bfm,th98-60}. 

In order to obtain information on the  CKM angle $\gamma$, in addition to
the ratio $R$ of the combined $B\to\pi K$ branching ratios introduced in 
(\ref{Def-R}), the ``pseudo-asymmetry'' 
\begin{equation}
A_0\equiv\frac{\mbox{BR}(B^0_d\to\pi^-K^+)-\mbox{BR}(\overline{B^0_d}\to
\pi^+K^-)}{\mbox{BR}(B^+\to\pi^+K^0)+\mbox{BR}(B^-\to\pi^-\overline{K^0})}=
A_{\rm CP}(B_d\to\pi^\mp K^\pm)\,R\,,
\end{equation}
as well as the quantities
\begin{equation}\label{r-eps-def}
r\equiv\frac{|T|}{\sqrt{\left\langle|P|^2\right\rangle}}\,,\quad
\epsilon\equiv\frac{|P_{\rm ew}|}{\sqrt{\left\langle|P|^2\right\rangle}}\,,
\end{equation}
where $\left\langle|P|^2\right\rangle$ is defined by
\begin{equation}
\left\langle|P|^2\right\rangle\equiv\frac{1}{2}\left(|P|^2+|\overline{P}|^2
\right)
\end{equation}
and
\begin{equation}\label{phases-def}
\delta\equiv\delta_T-\delta_{tc}\,,
\quad\Delta\equiv\delta_{\rm ew}-\delta_{tc}
\end{equation}
are differences of the relevant strong phases, turn out to be very 
useful \cite{th98-60}. The general expressions for $R$ and $A_0$ in terms 
of the parameters specified in (\ref{r-eps-def}) and (\ref{phases-def}) 
are quite complicated and are given explicitly in \cite{th98-60}, where 
further details can be found. Let us here just briefly discuss the basic 
ideas that are at the basis of the constraints on the CKM angle $\gamma$ 
arising from these observables.

The pseudo-asymmetry $A_0$ allows us to eliminate the strong phase $\delta$ 
in the expression for $R$. Consequently, if both $R$ and $A_0$ have been 
measured, contours in the $\gamma$--$r$ plane can be fixed. If the parameter 
$r$, i.e.\ $|T|$, could also be fixed, we were in a position to extract 
the value of $\gamma$ from these contours up to a four-fold ambiguity 
\cite{PAPIII,groro}. Unfortunately, since $T$ is not just a ``tree''
amplitude, $r$ may in general receive sizeable non-factorizable 
contributions. Therefore, expectations relying on ``factorization'' that 
a future theoretical accuracy of $r$ as small as ${\cal O}(10\%)$ may be 
achievable (see, for instance, \cite{groro,wuegai}) appear too optimistic. 

It is, however, in principle possible to {\it constrain} the CKM angle 
$\gamma$ in a way that does {\it not} depend on $r$, introducing the major 
theoretical uncertainty into the extraction of $\gamma$ sketched in the 
previous paragraph. Provided $R$ turns out to be smaller than 1, an 
interval around $\gamma=90^\circ$ can be ruled out \cite{fm2}, which is of 
particular phenomenological importance \cite{gnps}. As soon as a 
non-vanishing value of $A_0$ has been measured, also regions for $\gamma$ 
around $0^\circ$ and $180^\circ$ can be excluded. These constraints on 
$\gamma$ are related to the fact that $R$ (considered as a function of $r$; 
$\delta$ has been eliminated through $A_0$) takes a minimal value, which is 
given by \cite{th98-60}
\begin{equation}\label{Rmin}
R_{\rm min}=\kappa\,\sin^2\gamma\,+\,
\frac{1}{\kappa}\left(\frac{A_0}{2\,\sin\gamma}\right)^2,
\end{equation}
where 
\begin{equation}\label{kappa-def}
\kappa=\frac{1}{w^2}\left[\,1+2\,(\epsilon\,w)\cos\Delta+
(\epsilon\,w)^2\,\right]\quad\mbox{with}\quad
w=\sqrt{1+2\,\rho\,\cos\theta\cos\gamma+\rho^2}\,.
\end{equation}
The theoretical accuracy of the corresponding bounds on $\gamma$ is limited
both by rescattering processes and by contributions from electroweak 
penguins, which are included in $\kappa$ through the parameters $\rho$ and 
$\epsilon$, respectively. Neglecting these effects, we simply have $\kappa=1$,
which corresponds to the case discussed in the original paper \cite{fm2} on 
the bounds on $\gamma$ arising from $B\to\pi K$ decays. Let us focus in the 
following section on the rescattering processes, which have led to 
considerable interest in the recent literature \cite{gewe}--\cite{atso}. 
A detailed discussion of electroweak penguin effects can be found in 
\cite{th98-60} (see also \cite{neubert,groro}).

\boldmath
\section{Controlling the Rescattering Effects}\label{sec:FSIcontrol}
\unboldmath
The parameter $\rho$ describing the ``strength'' of the rescattering processes 
is highly CKM-suppressed by $\lambda^2R_b\approx0.02$, as can be seen in 
(\ref{rho-def}). Model calculations performed at the perturbative quark level 
give $\rho={\cal O}(1\%)$ and do not indicate a significant compensation of 
this very large CKM suppression. However, in a recent attempt \cite{fknp} 
to evaluate rescattering processes of the kind $B^+\to\{\pi^0K^+,\,
\pi^0K^{\ast +},\,\rho^0K^{\ast +},\,\ldots\,\}\to\pi^+K^0$,
it is found that $|{\cal P}_{uc}|/|{\cal P}_{tc}|={\cal O}(5)$,
implying that $\rho$ may be as large as ${\cal O}(10\%)$. A similar feature
arises also in a simple model to describe final-state interactions, which
has been proposed in \cite{gewe,neubert}. We will use this model for 
illustrative purposes in Section~\ref{sec:model}, where it is discussed in 
more detail.

Although it has been claimed by several authors in recent literature
that such rescattering processes would invalidate the constraints on the
CKM angle $\gamma$ implied by the $B_d\to\pi^\mp K^\pm$, $B^\pm\to\pi^\pm K$ 
observables, this is actually not the case \cite{th98-60}. These effects can 
be included in the bounds on $\gamma$ by using additional experimental data. 
The purpose of this section is to give a detailed discussion of this important 
feature, which will be illustrated in a quantitative way in the following 
section. 

A first step towards the control of rescattering processes is provided by 
the CP-violating asymmetry
\begin{equation}\label{Ap-def}
A_+\equiv\frac{\mbox{BR}(B^+\to\pi^+K^0)-\mbox{BR}(B^-\to\pi^-
\overline{K^0})}{\mbox{BR}(B^+\to\pi^+K^0)+\mbox{BR}(B^-\to\pi^-
\overline{K^0})}=-\,\frac{2\,\rho\,\sin\theta\sin\gamma}{1+
2\,\rho\,\cos\theta\cos\gamma+\rho^2}\,.
\end{equation}
While simple quark-level estimates give at most a few percent for this
CP asymmetry~\cite{pert-pens}, rescattering processes may lead to values 
as large as ${\cal O}(10\%)$ \cite{gewe}--\cite{atso}. As soon as $A_+$ 
has been measured, we are in a position to obtain upper and lower bounds 
on $\rho$, which are given by 
\begin{equation}\label{rho-min-max}
\rho_{\rm min}^{\rm max}=\frac{\sqrt{A_+^2\,+\,
\left(1-A_+^2\right)\sin^2\gamma}\,\pm\,\sqrt{\left(1-A_+^2\right)
\sin^2\gamma}}{|A_+|}\,.
\end{equation}
In particular the lower bound $\rho_{\rm min}$ is of special interest. A
detailed study can be found in \cite{th98-60}. In order to go beyond these
constraints, the decay $B^+\to K^+\overline{K^0}$ and its charge conjugate --
the $SU(3)$ counterparts of $B^\pm\to\pi^\pm K$ -- play a key role. 
The corresponding decay amplitude takes the form
\begin{equation}
A(B^+\to K^+\overline{K^0})=\lambda^3A\left[1-\left(
\frac{1-\lambda^2}{\lambda^2}\right)
\rho^{(d)}\,e^{i\theta_d}\,e^{i\gamma}\right]{\cal P}_{tc}^{(d)}\,,
\end{equation}
where
\begin{equation}\label{rhod-def}
\rho^{(d)}\,e^{i\theta_d}=\frac{\lambda^2R_b}{1-\lambda^2/2}
\left[1-
\left(\frac{{\cal P}_{uc}^{(d)}+{\cal A}^{(d)}}{{\cal P}_{tc}^{(d)}}
\right)\right]
\end{equation}
corresponds to (\ref{rho-def}), and direct CP violation is described by
\begin{eqnarray}
\lefteqn{A_+^{(d)}\equiv\frac{\mbox{BR}(B^+\to K^+\overline{K^0})-
\mbox{BR}(B^-\to K^-K^0)}{\mbox{BR}(B^+\to K^+\overline{K^0})+
\mbox{BR}(B^-\to K^-K^0)}}\nonumber\\
&&~~~=\frac{2\,\lambda^2\,(1-\lambda^2)\,\rho^{(d)}
\sin\theta_d\,\sin\gamma}{\lambda^4-2\,\lambda^2\,(1-\lambda^2)\,\rho^{(d)}
\cos\theta_d\,\cos\gamma+(1-\lambda^2)^2\rho^{(d)\,2}}\,.\label{Apd-def}
\end{eqnarray}
Moreover, the following ratio of combined branching ratios turns out to
be very useful~\cite{bfm}:
\begin{eqnarray}
\lefteqn{H\equiv R_{SU(3)}^{\,2}\left(\frac{1-\lambda^2}{\lambda^2}\right)
\frac{\mbox{BR}(B^\pm\to K^\pm K)}{\mbox{BR}(B^\pm\to\pi^\pm K)}}\nonumber\\
&&~=\frac{\lambda^4-2\,\lambda^2\,(1-\lambda^2)\,\rho^{(d)}\cos\theta_d\,
\cos\gamma+(1-\lambda^2)^2\rho^{(d)\,2}}{\lambda^4\left(1+
2\,\rho\,\cos\theta\,\cos\gamma+\rho^2\right)}\,.\label{H-def}
\end{eqnarray}
Here BR$(B^\pm\to K^\pm K)$ is defined in analogy to (\ref{BR-char}),
tiny phase-space effects have been neglected (for a more detailed 
discussion, see \cite{fm2}), and 
\begin{equation}
R_{SU(3)}=\frac{M_B^2-M_\pi^2}{M_B^2-M_K^2}\,
\frac{F_{B\pi}(M_K^2;0^+)}{F_{BK}(M_K^2;0^+)}
\end{equation}
describes factorizable $SU(3)$ breaking. Using the model of Bauer, 
Stech and Wirbel \cite{BSW} to estimate the relevant form factors, we have 
$R_{SU(3)}={\cal O}(0.7)$. At present, there is unfortunately no reliable 
approach available to deal with non-factorizable $SU(3)$ breaking.

If we look at (\ref{Ap-def}), (\ref{Apd-def}) and (\ref{H-def}), we observe 
that $A_+$, $A_+^{(d)}$ and $H$ depend on the four ``unknowns'' $\rho$, 
$\theta$, $\rho^{(d)}$, $\theta_d$, and of course also on the CKM angle 
$\gamma$. A possible strategy is to use
\begin{equation}\label{SU3-input}
\rho=\zeta_\rho\,\rho^{(d)}\,,
\end{equation}
where $\zeta_\rho$ parametrizes $SU(3)$-breaking corrections, in order to 
express $\rho^{(d)}$ in (\ref{Apd-def}) and (\ref{H-def}) through $\rho$. 
As a first ``guess'', we may use $\zeta_\rho=1$. The strong phase 
$\theta_d$ can be eliminated in $H$ with the help of the CP asymmetry 
$A_+^{(d)}$ arising in $B^\pm\to K^\pm K$, while $\theta$ can be eliminated 
through the CP asymmetry $A_+$ arising in $B^\pm\to\pi^\pm K$. Following these 
lines, we arrive at an expression for $H$, which depends only on $\rho$, 
$\gamma$, and on the $SU(3)$-breaking parameter $\zeta_\rho$. Consequently,
specifying $R_{SU(3)}$ and $\zeta_\rho$, for instance through $R_{SU(3)}=0.7$ 
and $\zeta_\rho=1$, we are in a position to determine $\rho$ and $\theta$ as 
functions of $\gamma$. In order to include the rescattering effects in the 
contours in the $\gamma$--$r$ plane and the bounds on $\gamma$ arising from 
(\ref{Rmin}), $\rho$ and $\theta$ determined this way are sufficient 
\cite{th98-60}. Keeping the $SU(3)$-breaking parameters $R_{SU(3)}$ and 
$\zeta_\rho$ explicitly in the corresponding formulae, it is possible to 
study the sensitivity to their chosen values, and to take into account 
$SU(3)$ breaking once we have a better understanding of this phenomenon. 
 
To simplify the following discussion, let us assume 
\begin{equation}\label{SU3-input2}
\rho=\rho^{(d)}\,\,\quad\mbox{and}\quad\,\,\theta=\theta_d\,.
\end{equation}
As has already been pointed out in \cite{bfm,th98-60}, this $SU(3)$ input
implies a nice relation between $A_+$, $A_+^{(d)}$ and the combined 
$B^\pm\to\pi^\pm K$ and $B^\pm\to K^\pm K$ branching ratios, which is given by 
\begin{equation}\label{ApApdrel}
\frac{A_+}{A_+^{(d)}}=-\,R_{SU(3)}^{\,2}\,\frac{\mbox{BR}(B^\pm\to K^\pm 
K)}{\mbox{BR}(B^\pm\to\pi^\pm K)}=-\left(\frac{\lambda^2}{1-\lambda^2}
\right)H\,,
\end{equation}
and allows the determination of $H$ and of the $SU(3)$-breaking parameter 
$R_{SU(3)}$ directly from the measured $B^\pm\to\pi^\pm K$ and 
$B^\pm\to K^\pm K$ observables. Using (\ref{H-def}) and (\ref{SU3-input2}), 
it is an easy exercise to derive the expression
\begin{equation}\label{expr}
2\,\rho\,\cos\theta\,\cos\gamma=a\,+\,b\,\rho^2
\end{equation}
with
\begin{equation}
a=\lambda^2\left[\frac{1-H}{1+\lambda^2\left(H-1\right)}\right],\quad
b=\frac{1}{\lambda^2}\left[\frac{\left(1-\lambda^2\right)^2-
\lambda^4\,H}{1+\lambda^2\left(H-1\right)}\right],
\end{equation}
which leads to
\begin{equation}\label{w-deter}
w=\frac{1}{\lambda}\,\sqrt{\frac{\rho^2+\lambda^2\left(1-\rho^2\right)}{1+
\lambda^2\left(H-1\right)}}\,.
\end{equation}
Combining (\ref{expr}) with the CP-violating asymmetry (\ref{Ap-def}), we 
obtain a quadratic equation for $\rho^2$, which has the solution
\begin{equation}\label{rho-constr}
\rho^2=\frac{w\,\pm\,\sqrt{w^2\,-\,u\,v}}{v}\,,
\end{equation}
where
\begin{eqnarray}
u&=&\left[\,a\,\sin^2\gamma\,+\,\left(a+1\right)A_+^2\cos^2\gamma\,\right]^2+
\,\left(\,A_+\,\sin\gamma\,\cos\gamma\,\right)^2\\
v&=&\left[\,b\,\sin^2\gamma\,+\,\left(b+1\right)A_+^2\cos^2\gamma\,\right]^2+
\,\left(\,A_+\,\sin\gamma\,\cos\gamma\,\right)^2\\
w&=&\left[\,A_+^2\,+\,2\,\left(1-A_+^2\right)\sin^2\gamma\,\right]
\sin^2\gamma\,\cos^2\gamma\,-\nonumber\\
&&\left[\,a\,\sin^2\gamma\,+\,\left(a+1\right)A_+^2\cos^2\gamma\,\right]
\left[\,b\,\sin^2\gamma\,+\,\left(b+1\right)A_+^2\cos^2\gamma\,\right].
\end{eqnarray}

The present upper limit on the combined $B^\pm\to K^\pm K$ branching ratio 
obtained by the CLEO collaboration is given by $2.1\times10^{-5}$ \cite{cleo}.
At first sight, experimental studies of this mode appear to be difficult, 
since the ``short-distance'' expectation for its combined branching ratio 
is ${\cal O}(10^{-6})$ (see, for instance, \cite{pert-pens}). However, as 
was pointed out in \cite{th98-60}, rescattering effects may enhance this 
observable by a factor as large as ${\cal O}(10)$, and could thereby 
make $B^\pm\to K^\pm K$ measurable at future $B$ factories. In the following 
section, we illustrate this feature, as well as the constraints on $\gamma$ 
and the strategy to control the rescattering processes affecting them, in
a quantitative way, by using a simple model.

\boldmath
\section{An Illustration within a Simple Model}\label{sec:model}
\unboldmath
In Ref.~\cite{gewe}, a simple model was introduced to deal with final-state 
interactions in $B\to\pi K$ decays, which has been refined in \cite{neubert}, 
where also electroweak penguin effects are considered. The basic idea of 
this model is very simple: the generalized factorization prescription 
\cite{ns} is used to calculate the ``short-distance'' contributions to the  
$B\to\pi K$ decay amplitudes, whereas the ``long-distance'' effects are 
taken into account by simply introducing {\it elastic} rescattering phases 
$\phi_{1/2}$ and $\phi_{3/2}$ for the two isospin channels of the 
final-state mesons. Following these lines, we obtain 
\begin{eqnarray}
A(B^+\to\pi^+K^0)&=&\left[\,e^{i\phi_P}\,+\,
\frac{1}{3}\,z\left\{1+\left(1+\frac{1}{y_{\rm ew}}
\right)\left(e^{i\Delta\phi}-1\right)\right\}\right.\nonumber\\
&&\left.-\,\frac{1}{3}\,x\,(1+y)
\left(e^{i\Delta\phi}-1\right)\,e^{i\gamma}\right]|M_P|\,
e^{i\phi_{1/2}}\label{Ap-elastic}
\end{eqnarray}
\begin{eqnarray}
A(B^0_d\to\pi^-K^+)&=&-\left[\,e^{i\phi_P}\,-\,
\frac{1}{3}\,z\left\{2+\left(1+\frac{1}{y_{\rm ew}}
\right)\left(e^{i\Delta\phi}-1\right)\right\}\right.\nonumber\\
&&\left.+\,x\left\{1+\frac{1}{3}\,(1+y)\left(e^{i\Delta\phi}-1\right)
\right\}\,e^{i\gamma}\right]|M_P|\,e^{i\phi_{1/2}},\label{An-elastic}
\end{eqnarray}
where $\Delta\phi=\phi_{3/2}-\phi_{1/2}$ is the difference of the elastic
rescattering phases, and $\phi_P$ has been introduced to describe the
strong phase of penguin topologies with internal charm quarks, which receive
important contributions from rescattering processes of the kind $B^+\to
\{\overline{D^0}D_s^+,\,\overline{D^0}D_s^{\ast+},\,
\overline{D^{\ast 0}}D_s^{\ast+},\,\ldots\,\}\to\pi^+K^0$. Neglecting such 
effects, $\phi_P$ takes the trivial value $180^\circ$, which is related to 
the minus sign appearing in (\ref{P-def}). In this simple model, final-state
interactions can be  ``switched on'' by choosing a non-vanishing value for
$\Delta\phi$. If we denote the colour-allowed and colour-suppressed ``tree'' 
amplitudes obtained within the framework of generalized factorization 
\cite{neubert,ns} by $M_T$ and $M_C$, respectively, and the corresponding 
$\bar b\to\bar s$ QCD penguin amplitude by $M_P$, the parameters $x$ and $y$
are given by 
\begin{equation}
x=\frac{|M_T|}{|M_P|}\approx0.2\,,\quad
y=\frac{|M_C|}{|M_{T}|}\approx
\frac{a_2}{a_1}\approx0.25\,,
\end{equation}
where $a_1$ and $a_2$ denote the usual phenomenological colour factors. 
The origin of the parameters $z$ and $y_{\rm ew}$ is due to electroweak 
penguins. They are given by
\begin{equation}
z=\frac{|M_{\rm ew}^{\rm C}|}{|M_P|}\,,\quad y_{\rm ew}=
\frac{|M_{\rm ew}^{\rm C}|}{|M_{\rm ew}|}\,,
\end{equation}
where $M_{\rm ew}^{\rm C}$ and $M_{\rm ew}$ are the colour-suppressed and
colour-allowed electroweak penguin amplitudes, again calculated by using 
generalized factorization. We have $z/(x\,y)\approx 0.75$ and 
$y_{\rm ew}\approx y$, where the derivation of the numerical factor 
0.75 can be found in \cite{th98-60}. Using ({\ref{Ap-elastic}) and 
(\ref{An-elastic}), it is an easy exercise to calculate $\rho$, $T$ and 
$P_{\rm ew}$ in our simple model:
\begin{eqnarray}
\rho\,e^{i\theta}&=&-\,\frac{x\,(1+y)\left(e^{i\Delta\phi}-1\right)}{3\,
e^{i\phi_P}\,+\,z\left[1+\left(1+1/y_{\rm ew}\right)
\left(e^{i\Delta\phi}-1\right)\right]}\label{rho-elastic}\\
|T|\,e^{i\delta_T}&=&x\left[1+\frac{2}{3}\left(1+y\right)
\left(e^{i\Delta\phi}-1\right)\right]|M_P|\,e^{i\phi_{1/2}}\label{T-elastic}\\
|P_{\rm ew}|\,e^{i\delta_{\rm ew}}&=&z\left[1+\frac{2}{3}
\left(1+\frac{1}{y_{\rm ew}}\right)\left(e^{i\Delta\phi}-1\right)
\right]|M_P|\,e^{i\phi_{1/2}}\,,\label{Pew-elastic}
\end{eqnarray}
and correspondingly $r$ and $\epsilon$, which are obtained by normalizing 
$|T|$ and $|P_{\rm ew}|$ through $\sqrt{\langle|P|^2\rangle}$ 
(see (\ref{r-eps-def})).

The formulae given in \cite{th98-60} allow us to calculate all relevant 
observables of the $B^\pm\to\pi^\pm K$ and $B_d\to\pi^\mp K^\pm$ decays. 
In Fig.~\ref{fig:Rmin}, we show the resulting dependence of the ratio
$R$ of combined $B\to\pi K$ branching ratios (\ref{Def-R}) on the CKM angle 
$\gamma$ for $x=0.2$, $y=0.25$, $z=0.0375$, $\phi_P=180^\circ$ and various 
values of $\Delta\phi$. In this figure, we have also included the curves 
corresponding to $R_{\rm min}$ (see (\ref{Rmin})) in order to illustrate 
the way in which the corresponding bounds on $\gamma$ are realized in this 
specific example. Note that the difference between the thin and thick solid 
lines is due to electroweak penguins.

Concerning the CP asymmetries $A_0$ and $A_+$, we ``naturally'' get values
as large as ${\cal O}(10\%)$. An interesting relation arises between these
observables for $\phi_P=180^\circ$. If we neglect the electroweak penguin 
contributions for a moment, i.e.\ $z=0$, we obtain
\begin{eqnarray}
A_0&=&\frac{-\,6\,x\left[(1+y)\sin(\Delta\phi-\phi_P)-(2-y)\sin\phi_P\right]
\sin\gamma}{9+6\,x\,(1+y)\left[\cos\phi_P-
\cos(\Delta\phi-\phi_P)\right]\cos\gamma+2\,x^2(1+y)^2(1-\cos\Delta\phi)}\\
A_+&=&\frac{6\,x\,(1+y)\left[\,\sin(\Delta\phi-\phi_P)+\sin\phi_P\right]
\sin\gamma}{9+6\,x\,(1+y)\left[\cos\phi_P-\cos(\Delta\phi-\phi_P)\right]
\cos\gamma+2\,x^2(1+y)^2(1-\cos\Delta\phi)}\,.
\end{eqnarray}
Consequently, in the case of $\phi_P=180^\circ$, we have $A_0=-\,A_+$. This
relation is only affected to a small extent by electroweak penguins. Since
the CP asymmetries are, however, very sensitive to the strong phase $\phi_P$, 
it may easily be spoiled through $\phi_P\not=180^\circ$. 

\begin{figure}
\centerline{
\rotate[r]{
\epsfxsize=9.2truecm
\epsffile{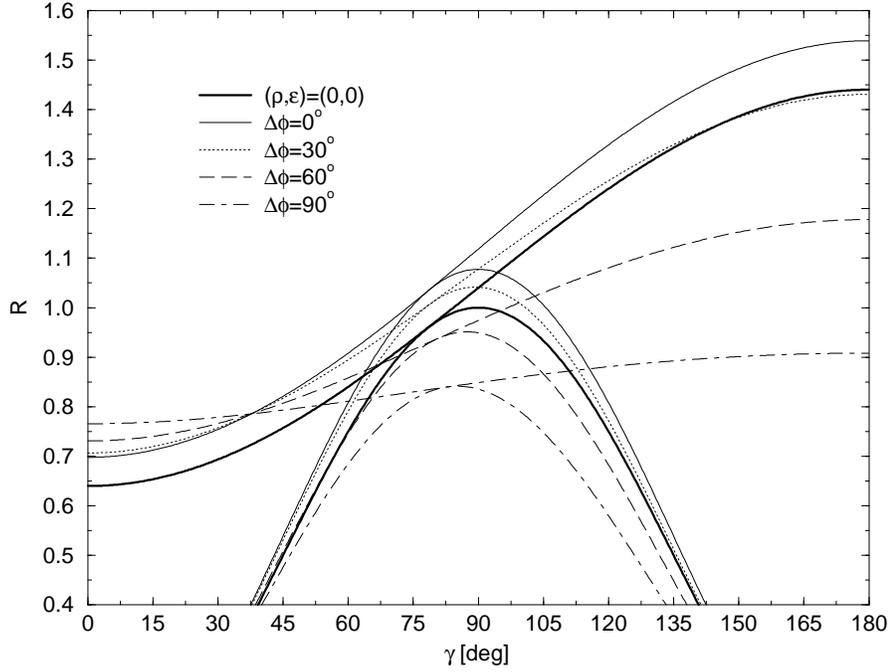}}}
\caption{Illustration of the bounds on the CKM angle $\gamma$ arising
from (\ref{Rmin}) within a simple model of final-state interactions specified
in the text.}\label{fig:Rmin}
\end{figure}

\begin{figure}
\centerline{
\rotate[r]{
\epsfxsize=9.2truecm
\epsffile{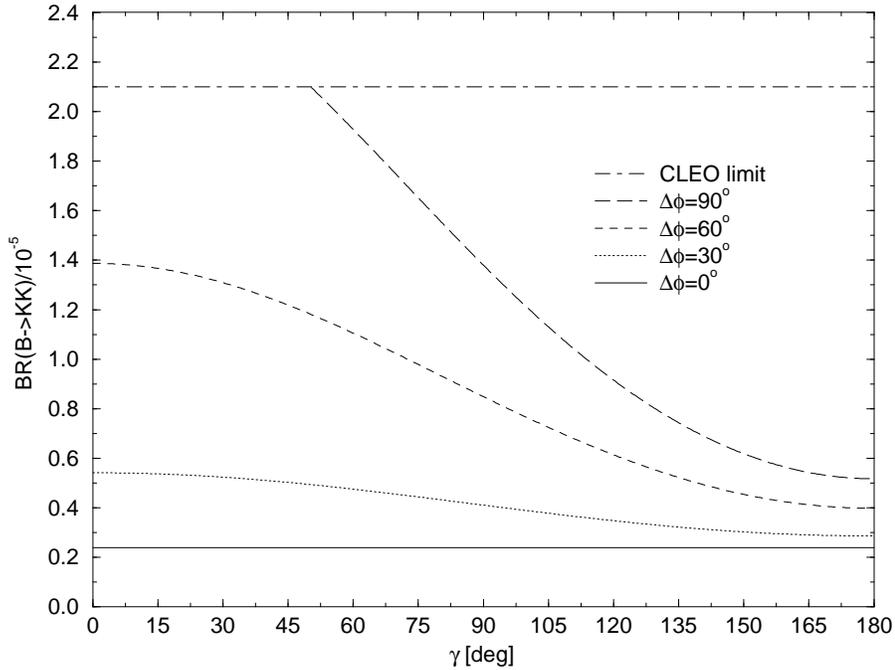}}}
\caption{The dependence of the combined branching ratio BR$(B^\pm\to K^\pm K)$
on the CKM angle $\gamma$ for a simple model of final-state interactions 
specified in the text.}\label{fig:BRbkk}
\end{figure}

\begin{figure}
\centerline{
\rotate[r]{
\epsfxsize=9.2truecm
\epsffile{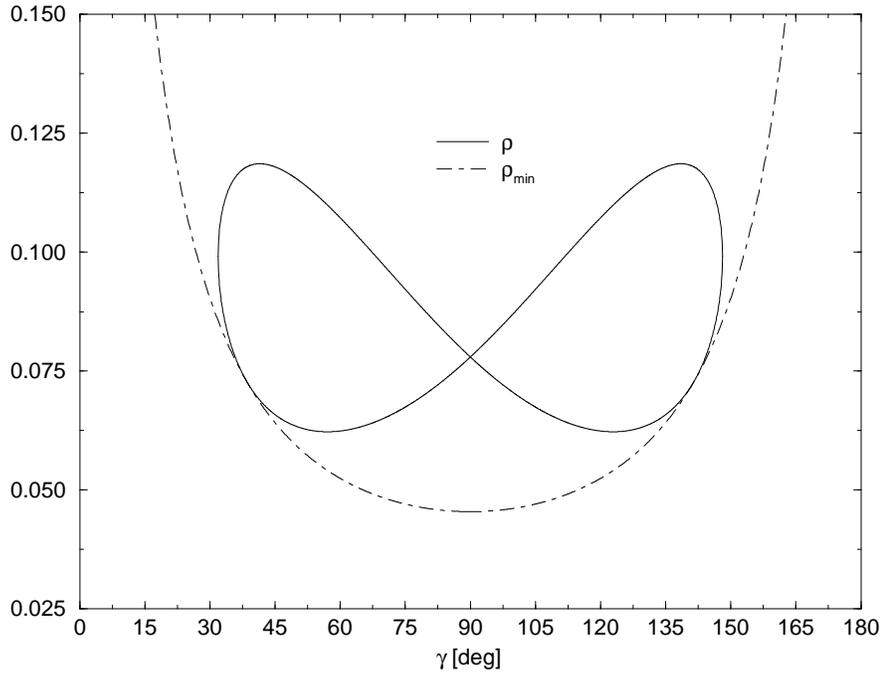}}}
\caption{The dependence of $\rho$ determined with the help of 
(\ref{rho-constr}) on the CKM angle $\gamma$ for a specific example 
discussed in the text.}\label{fig:rho}
\end{figure}

\begin{figure}
\centerline{
\rotate[r]{
\epsfxsize=9.2truecm
\epsffile{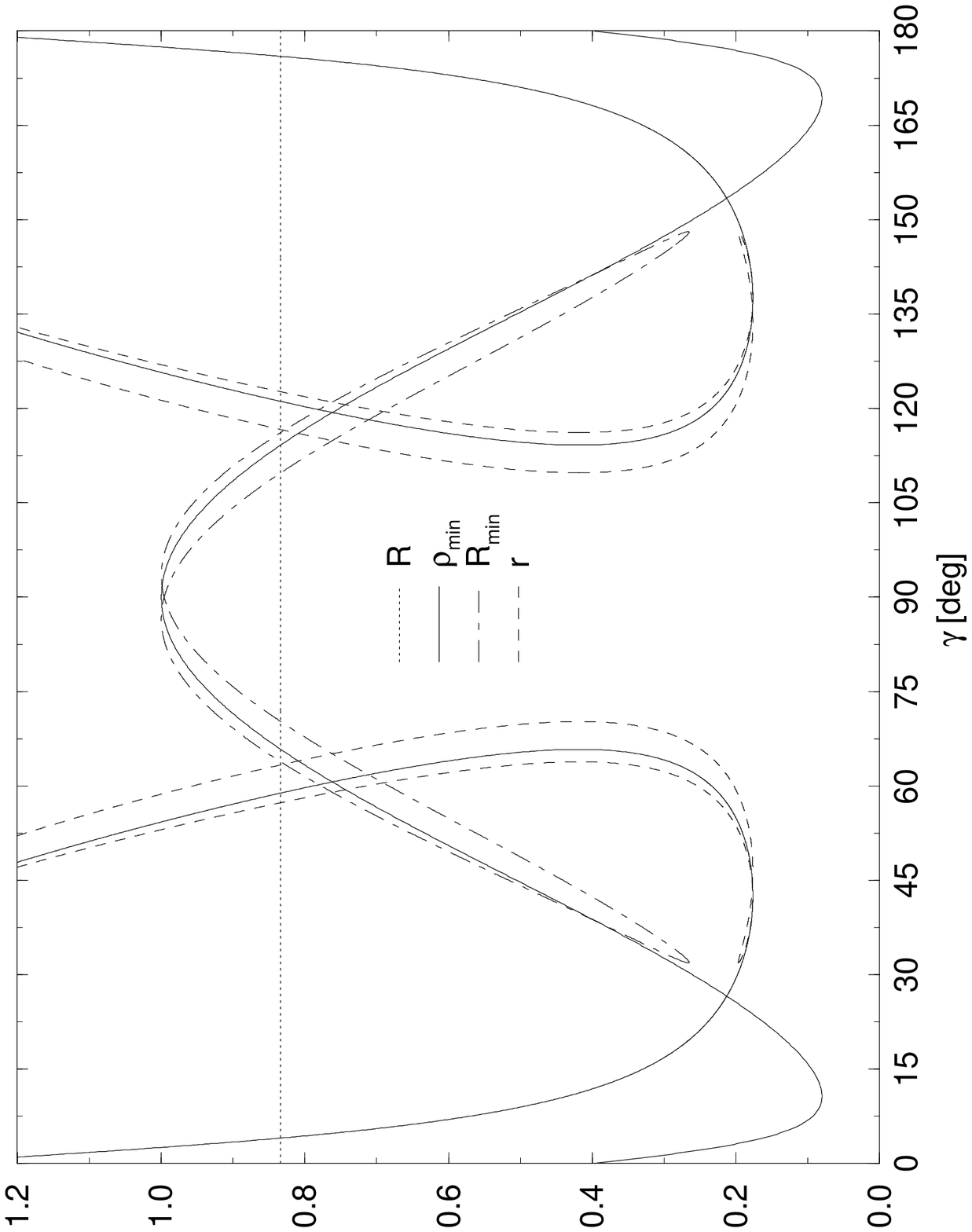}}}
\caption{Controlling rescattering effects in $R_{\rm min}$ and the contours
in the $\gamma$--$r$ plane through the $B^\pm\to K^\pm K$ observables for a 
specific example discussed in the text.}\label{fig:FSIcontrol}
\end{figure}

At the end of the previous section we noted that the combined 
branching ratio BR$(B^\pm\to K^\pm K)$ may be enhanced significantly 
through final-state interaction effects. This interesting feature can 
be seen nicely in Fig.~\ref{fig:BRbkk}, where we have used the same 
input parameters $x$, $y$, $z$ and $\phi_P$ as above, $R_{SU(3)}=0.7$, 
the $SU(3)$ relation (\ref{SU3-input2}), and 
BR$(B^\pm\to\pi^\pm K)=2.3\times10^{-5}$. The CP asymmetry $A_+^{(d)}$,
which depends strongly on the CKM angle $\gamma$, may well be as large 
as ${\cal O}(50\%)$.

Let us now have a closer look at the strategy to control the rescattering 
processes discussed in Section~\ref{sec:FSIcontrol}. To this end, we choose
$\gamma=50^\circ$, $\Delta\phi=45^\circ$, $\phi_P=180^\circ$, and the same 
values for $x$, $y$ and $z$ as in our previous examples. Then we obtain 
$R=0.83$, $A_+=-\,9.1\%$, $A_0=8.0\%$, BR$(B^\pm\to K^\pm K)=7.9\times10^{-6}$,
and $A_+^{(d)}=54\%$. For the $B^\pm\to K^\pm K$ observables, we have assumed 
in addition BR$(B^\pm\to\pi^\pm K)=2.3\times10^{-5}$, $R_{SU(3)}=0.7$, and 
the $SU(3)$ relation (\ref{SU3-input2}). Supposing that a future $B$-factory
experiment will find these values, the ratio (\ref{ApApdrel}) would imply 
$H=3.3$ and $R_{SU(3)}=0.7$. Using now (\ref{rho-constr}), $\rho$ can be 
constrained as shown in Fig.~\ref{fig:rho}, implying $\rho=0.090\pm0.028$. 
The ``true'' value of $\rho$ is given in this example by 0.063 and is very 
close to its lower bound. The dot-dashed line in Fig.~\ref{fig:rho} 
corresponds to the ``minimal'' value $\rho_{\rm min}$, which can be obtained 
from $A_+$ with the help of (\ref{rho-min-max}). The non-vanishing direct 
CP asymmetries also imply a range for $\gamma$, which is given by 
$32^\circ\leq\gamma\leq148^\circ$, and excludes values around $0^\circ$ and 
$180^\circ$. 

Using (\ref{rho-constr}}), the rescattering effects can be included in 
$R_{\rm min}$ through (\ref{kappa-def}). The corresponding curves are 
represented in Fig.~\ref{fig:FSIcontrol} by the dot-dashed lines, whereas 
the dotted line corresponds to the ``measured'' value $R=0.83$, which is 
larger than the present central value 0.65, and excludes the range 
$70^\circ\leq\gamma \leq110^\circ$ around $90^\circ$. The contours 
in the $\gamma$--$r$ plane, taking into account the rescattering effects, 
are represented by the dashed lines in Fig.~\ref{fig:FSIcontrol}. As in 
Fig.~\ref{fig:rho}, we have also included the curves corresponding to 
$\rho_{\rm min}$ (the solid lines), which can be constructed by using only 
the $B\to\pi K$ observables, i.e.\ without making use of $B^\pm \to K^\pm K$. 
Consequently, in our example, the allowed range for $\gamma$ would be given 
by $32^\circ\leq\gamma\leq70^\circ$~$\lor$~$110^\circ\leq\gamma\leq148^\circ$,
while the ``true'' value is $\gamma=50^\circ$. It is interesting to note 
that additional information on $r$ -- in our example, the ``true'' value is 
0.19 -- would not lead to a significantly more stringent range for $\gamma$ 
in this case, as can be seen in Fig.~\ref{fig:FSIcontrol}. 

Although electroweak penguins are included in our simple model and the 
values of the $B\to\pi K$ and $B^\pm\to K^\pm K$ observables calculated 
in this section, they are not included in the curves shown in 
Fig.~\ref{fig:FSIcontrol}. In the case of our example, we have $\epsilon=
0.089$ and $\Delta=92^\circ$. Since $\Delta$ is very close to $90^\circ$,
the electroweak penguin effects in the bound on $\gamma$ are only of second 
order in $\epsilon$, as can be seen in (\ref{kappa-def}). Consequently,
despite the large value of $\epsilon$ -- the ``short-distance'' value is
$\epsilon={\cal O}(0.03)$ -- electroweak penguins affect the constraints 
on $\gamma$ only to a small extent in our example. In general, however, 
electroweak penguin effects may represent the most important limitation of
the theoretical accuracy of the bounds on $\gamma$. A detailed analysis
can be found in \cite{th98-60}.

\boldmath
\section{Conclusions}\label{sec:sum}
\unboldmath
We have shown that the decay $B^+\to K^+\overline{K^0}$ and its charge
conjugate allow us to take into account rescattering effects in constraints
on the CKM angle $\gamma$ arising from $B\to\pi K$ modes. To accomplish
this task, the $SU(3)$ flavour symmetry of strong interactions has to be 
used in order to relate $B^\pm\to K^\pm K$ to $B^\pm\to\pi^\pm K$. An 
important by-product of this approach is an allowed range for the parameter 
$\rho$, measuring the strength of the rescattering processes. Concerning 
the bounds on $\gamma$, $SU(3)$ breaking enters only at the ``next-to-leading 
order'' level, as it represents a correction to the correction that is due 
to the rescattering processes. Moreover, we have also indicated ways to 
explore the impact of $SU(3)$ breaking in a quantitative way. 

In order to illustrate this strategy and the constraints on $\gamma$, we 
have used a simple model to describe final-state interactions. A realistic 
description is unfortunately out of reach at present, and would have to 
include, for instance, also inelastic rescattering contributions, which 
are expected to play an important role \cite{inel}, and are neglected in 
this model. Other shortcomings are, for instance, the question of 
whether $a_1$ and $a_2$ take values similar to those measured in 
$B\to D^{(\ast)}\,\pi\,(\rho)$ decays, or the problem to fix 
the ``short-distance'' QCD penguin amplitude $|M_P|$. This model can
therefore only serve to illustrate certain qualitative features of 
rescattering effects, for example their tendency to enhance the combined 
$B^\pm\to K^\pm K$ branching ratio significantly, or to induce sizeable 
CP violation in $B^\pm\to\pi^\pm K$.

In Refs.~\cite{gewe,neubert}, this model has been used to ``demonstrate'' 
that no useful constraints on the CKM angle $\gamma$ can be obtained from 
$B\to\pi K$ decays in the presence of final-state interactions (similar 
statements, although somewhat more moderate, have been made in 
\cite{fknp,atso} within a different framework). Here we have shown -- using 
exactly this model -- that this is actually not the case. To this end, 
we have even chosen rather large values of $\Delta\phi$ in our examples, 
corresponding to large final-state interaction effects. In a recent attempt 
to calculate this strong phase by using a Regge pole model for $\pi\,K$ 
scattering, significantly smaller values, lying within the range between 
$14^\circ$ and $20^\circ$, have been obtained \cite{dgpw}. In that case, 
there would not even be the need to correct at all for the final-state 
interaction effects in the bounds on $\gamma$. Certainly, future experimental 
data will tell us how important final-state interactions in $B\to\pi K$ 
decays really are. 

\vspace{0.7cm}

\noindent
{\it Acknowledgement}

\vspace{0.3cm}

\noindent
I am grateful to Peter Gaidarev for e-mail correspondence concerning the 
experimental status of $B^\pm\to K^\pm K$.

\end{document}